\documentclass[prd,aps, reprint, nolongbibliography,twocolumn,amsfonts,amssymb,amsmath,showpacs,letterpaper,superscriptaddress,nofootinbib,altaffilletter]{revtex4-1}

\usepackage[citecolor=blue,linkcolor=blue,colorlinks=true,urlcolor=blue]{hyperref}

\usepackage{graphicx}
\usepackage{amssymb}
\usepackage{amsmath}
\usepackage{longtable}
\usepackage{rotating}
\usepackage{color}
\usepackage{epsf}
\usepackage{fancyhdr}
\usepackage{ifthen}
\usepackage{slashed}
\usepackage{xspace}
\usepackage{tabularx}

\newcommand{\Mcwb}{$M_{\mathrm{cWB}}$\xspace}
\newcommand{\plpi}{PLP$_{65}$\xspace}
\newcommand{\bpli}{BPL$_{65}$\xspace}
\newcommand{\plpii}{PLP$_{50}$\xspace}
\newcommand{\bplii}{BPL$_{50}$\xspace}
\newcommand{\plpiii}{PLP$_{40}$\xspace}
\newcommand{\bpliii}{BPL$_{40}$\xspace}
\newcommand{\Mone}{$m_{1}$\xspace}

\begin{document}

\title{Detection of LIGO-Virgo binary black holes in the pair-instability mass gap}

\author{B.~O'Brien}
\author{M.~Szczepa\'nczyk}
\author{V.~Gayathri}
\author{I.~Bartos}
\affiliation{Department of Physics, University of Florida, PO Box 118440, Gainesville, FL 32611-8440, USA}
\author{G.~Vedovato}
\address {Universit\`a di Padova, Dipartimento di Fisica e Astronomia, I-35131 Padova, Italy }
\address {INFN, Sezione di Padova, I-35131 Padova, Italy }
\author{G.~Prodi}
\affiliation{Universit\`a di Trento, Dipartimento di Fisica, I-38123 Povo, Trento, Italy}
\affiliation{INFN, Trento Institute for Fundamental Physics and Applications, I-38123 Povo, Trento, Italy}
\author{G.~Mitselmakher}
\author{S.~Klimenko}
\affiliation{Department of Physics, University of Florida, PO Box 118440, Gainesville, FL 32611-8440, USA}

\begin{abstract}

By probing the population of binary black hole (BBH) mergers detected by LIGO-Virgo, we can infer properties about the underlying black hole formation channels. A mechanism known as pair-instability (PI) supernova is expected to prevent the formation of black holes from stellar collapse with mass greater than $\sim 40-65\,M_\odot$ and less than $\sim 120\,M_\odot$. Any BBH merger detected by LIGO-Virgo with a component black hole in this gap, known as the PI mass gap, likely originated from an alternative formation channel. 
Here, we firmly establish GW190521 as an outlier to the stellar-mass BBH population if the PI mass gap begins at or below $65\, M_{\odot}$.
In addition, for a PI lower boundary of $40-50\, M_{\odot}$, we find it unlikely that the remaining distribution of detected BBH events, excluding GW190521, is consistent with the stellar-mass population.

\end{abstract}

\date[\relax]{Dated: \today }

\maketitle

\section{Introduction}

During the first two observing runs of Advanced LIGO and Advanced Virgo (O1 and O2), 11 gravitational wave (GW) signals were identified by the LIGO-Virgo Collaboration~\cite{aLIGO:2015, aVirgo:2014} including 10 from binary black hole (BBH) mergers~\cite{GWTC1}.
Substantial improvements in detector sensitivity prior to the first half of the third observing run (O3a) led to the detection of 39 new GW event candidates, at least 36 of which are consistent with BBH mergers~\cite{GWTC2}.
This growing population of BBH mergers allows us to probe underlying black hole formation channels~\cite{POP_GWTC1, POP_GWTC2}.

Black holes can form directly through stellar collapse, mergers, accretion, or through collapse of dense gas in the early Universe (primordial black holes).
In the case of stellar evolution, a mechanism known as pair-instability (PI) supernovae is expected to prevent the formation of heavier black holes~\cite{Woosley:2007ppi, Spera:2017fyx, Woosley:2017, Giacobbo:2017, Belczynski:2016jno}.
In PI, the production of positron-electron pairs causes a reduction in radiation pressure. This in turn causes the core of the star to contract, raising the temperature and increasing the production of positron-electron pairs in a runaway process.
Any star with a helium core mass $32\,M_{\odot} \lesssim M_{\mathrm{HE}} \lesssim 64\,M_{\odot}$ undergoes pulsational pair-instability (PPI). In PPI, the stellar envelope is removed and the mass of the star is reduced until the star reaches a stable state~\cite{Woosley:2007ppi,Barkat:1967,Chen:2014,Yoshida:2016}.
Any star with a helium core mass $64\,M_{\odot} \lesssim M_{\mathrm{HE}} \lesssim 135\,M_{\odot}$ undergoes PI supernova which blasts the star apart, leaving no remnant compact object~\cite{Fowler:1964,Woosley:2007ppi, Heger:2003}.
More massive stars are predicted to collapse directly to black holes~\cite{Bond:1984, Ober:1983, Heger:2003}.
The PI process results in an absence of black holes in the mass spectrum known as the PI mass gap.
Some theoretical uncertainties about the lower boundary remain~\cite{Farmer:2020, Limongi:2018, Costa_gw190521_pisn_2020, Belczynski_pisn_uncertainty_2020}, however, in the following work we will consider the boundary to be below $65\,M_\odot$~\cite{Heger:2003, Belczynski:2016jno, Spera:2017fyx, Woosley:2017, Woosley:2019, Giacobbo:2017}.
The upper boundary is more certain and is believed to be $\sim 120\,M_\odot$~\cite{Belczynski:2016jno, Spera:2017fyx, Woosley:2017}.

Black holes can exceed the PI mass limit if they are formed from the previous merger of smaller black holes. This so-called hierarchical merger scenario can occur in dynamical encounters in dense stellar clusters such as galactic nuclei and globular clusters~\cite{Miller:2002, OLeary:2006, Giersz:2015, Mapelli:2016, Spera:2019, DiCarlo:2019, Bouffanais:2019, DiCarlo:2020, Gayathri_GW170817A_2020, Veske_heirarch_2020, Yang_2019, Kimball:2020qyd}, or through gas-capture in the disks of active galactic nuclei (AGNs)~\cite{Bartos:2017, McKernan:2018, McKernan:2012, McKernan:2014, Bellovary:2016, McKernan:2020, Yang_agn_disk_2019}. Alternatively, Black holes can exceed the PI mass limit by substantial accretion, which is possible for black holes in AGN disks~\cite{Yang_2020,Tagawa_2021}.

GW190521 is the heaviest BBH merger detected to date and marks the first direct detection of an intermediate mass black hole~\cite{GW190521_discovery}.
Ref.~\cite{GW190521_implications} estimates the probability for at least one of the component black holes to lie within $65-120\,M_\odot$ to be $99.0 \%$, assuming a binary system with quasi-circular orbit.
This probability increases when the analysis was repeated after taking into account the population of detected binary systems~\cite{POP_GWTC2}.
Hence, GW190521 is an appealing candidate for alternative formation scenarios -- a prospect which has already been explored in various papers~\cite{gayathri_gw190521:2020, deluca_gw190521:2020,  Fishbach_gw190521:2020, Safarzadeh_gw190521:2020, Romero_Shaw_gw190521:2020, Baxter_2021, Edelman_2021}.
We caution, however, that analyzing the features of a potential outlier BBH event outside the context of the entire population of detected events can be misleading (a notion previously examined by Fishbach \textit{et al} in Ref.~\cite{Fishbach_outliers:2020}).

Here, we investigate whether GW190521 or any other BBH merger candidate is an outlier to the stellar-mass population and thus formed via alternative formation scenarios; though distinguishing between separate alternative formation channels is outside the scope of this work.
We examine the BBH events detected by coherent WaveBurst (cWB), a model waveform independent pipeline used to search for GW signals~\cite{Klimenko:2008fu, Klimenko:2016}, and compare this sample of detected events against models of the stellar-mass BBH population.
In the method presented below, we add simulated BBH waveforms (injections) directly into the GW detector strain data and recover them with the cWB pipeline.
This allows us to directly account for detector noise when estimating the significance of potential outliers.
Ultimately, we seek to answer two fundamental questions: (i) accounting for statistical fluctuations, are there any outliers to the stellar-mass BBH population? and (ii) is the distribution of detected BBH events consistent with the stellar-mass BBH population?

This paper is organized as follows.
In Section~\ref{sec:method}, we familiarize the reader with cWB -- a pipeline that uses minimal assumptions to identify GW transient signals. Next, we specify the stellar-mass BBH population models and injections used in this study. Then, we define our test statistic and describe our statistical procedure used to classify outliers.
In Section~\ref{sec:results}, we present our results for various possible models of the stellar-mass BBH population.
Finally, in Section~\ref{sec:conclusions}, we state the conclusions of this study and briefly discuss our future work.

\section{Method}
\label{sec:method}

\subsection{Coherent WaveBurst}
\label{sec:cwb}

CWB is a search algorithm that identifies GW signals in the LIGO-Virgo detector data without directly using a waveform model~\cite{Klimenko:2008fu, Klimenko:2016}.
The signal detection process is performed in the time-frequency domain using the Wilson Daubechies Meyer wavelet transform~\cite{Necula:2012}. Here, the algorithm identifies wavelets inconsistent with detector noise and assembles them into clusters. If a cluster of wavelets is coincident in time and frequency across the detector network, then an event trigger is generated.
For each event trigger, summary statistics are calculated which describe the time-frequency evolution ($f_0$, $\Delta T_\mathrm{s}$, $\Delta F_\mathrm{s}$, $\mathcal{M}$, $F_\mathrm{M}$, $e_\mathrm{M}$), signal strength and coherence (SNR$_{\mathrm{net}}$, $c_\mathrm{c}$, $n_\mathrm{f}$), and the likelihood fit ($E_\mathrm{c}/L$). 
Additional statistics are used to estimate the number of cycles in the reconstructed waveform ($Q_0$, $Q_1$).
The summary statistics used in this study are defined in Appendix~\ref{sec:features}.

Through O1, O2, and O3a, cWB has contributed to the detection of 22 BBH merger candidates (displayed in Table~\ref{tab:events}).
Among these detections is GW190521, which cWB played a crucial role in identifying~\cite{GW190521_discovery, GW190521_cwb}.
In this work, we compare these BBH detections against predictions made by the stellar-mass population.
Simulated waveforms representative of the stellar-mass population are injected into the GW detector data and recovered with cWB.
We restrict our analysis to events detected by cWB with a false alarm rate (FAR) less than $ 1\,\mathrm{yr}^{-1}$ to maintain a high purity sample set.
Although the inclusion of Virgo generally improves source property reconstruction, we simplify our analysis by considering only the Hanford-Livingston detector network.

\subsection{Stellar-mass BBH population models}
\label{sec:models}

 \begin{figure*}[!tbh]
    \centering
    \includegraphics[width=\textwidth]{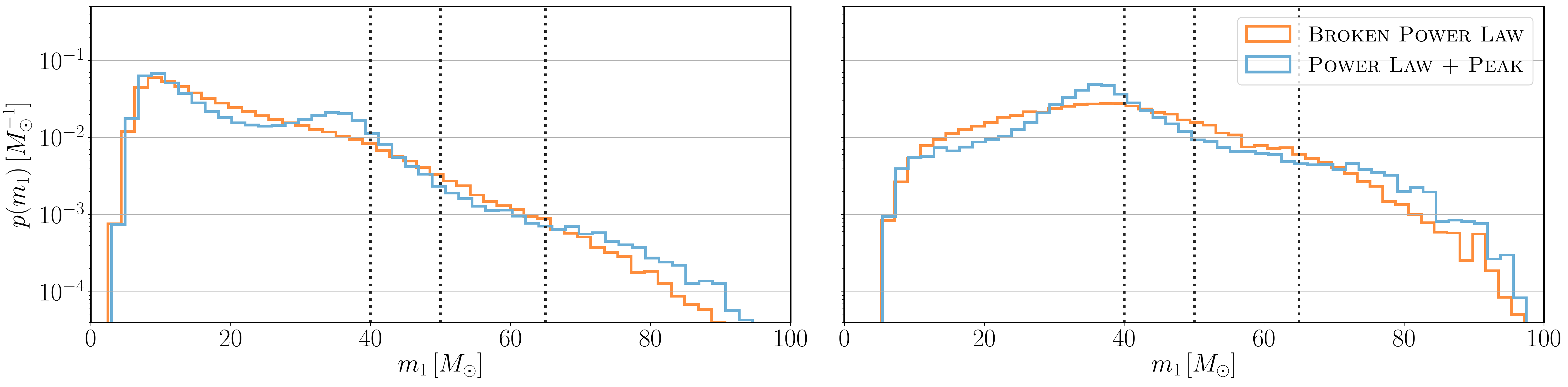}
    \caption{Expected probability distribution of source frame primary mass $m_1$, averaged over uncertainty, for the \textsc{Broken Power Law} (orange) and \textsc{Power Law  + Peak} (blue) models: (left) astrophysical distributions, (right) distributions of events detected by cWB with FAR less than $1\,\mathrm{yr}^{-1}$. Dotted vertical lines indicate tested values of the PI mass limit: $40\, M_{\odot}$, $50\, M_{\odot}$, and $65\, M_{\odot}$.}
    \label{fig:model_det_m1}
\end{figure*}

We examine two phenomenological mass models presented in Ref.~\cite{POP_GWTC2}: the \textsc{Broken Power Law} model~\cite{Broken_power_law} and the \textsc{Power Law  + Peak} model~\cite{Power_law_peak}.
Each population model is described by a set of hyper-parameters whose values are estimated by applying hierarchical Bayesian inference on the full population of BBH merger candidates detected by LIGO-Virgo~\cite{thrane_talbot:2019, Mandel:2019,vitale_inferring:2020}.
Below is a short description of each model: 

\begin{itemize}
  \item \textsc{Broken Power Law}: The primary mass distribution is characterized by a gentle sloping power law at low mass designed to capture the observed Salpeter initial mass function~\cite{Salpeter:1955,Salpeter_evidence} and a second steep sloping power law at high mass which targets a mass tapering of the distribution. The mass ratio is given as a power law that favors equal mass binaries~\cite{Belczynski:2020}.

  \item \textsc{Power Law  + Peak}: The primary mass distribution is characterized by a power law component at low mass and wide Gaussian component at high masses which models a build-up of black holes due to PPI supernovae~\cite{Heger:2002, Woosley:2015}. Similar to the previous model, the mass ratio is given as a power law favoring equal mass binaries.
\end{itemize}

The left plot in Figure~\ref{fig:model_det_m1} displays the astrophysical distribution for the primary mass \Mone predicted by each model, averaged over uncertainty.
These mass models describe the complete astrophysical distribution of BBHs and do not distinguish between separate formation channels.
To isolate the stellar-mass BBH population, we apply a theoretically motivated cutoff on the primary mass to model the PI mass gap lower boundary. We assume that no stellar-mass BBHs form inside of the PI gap.
To account for the uncertainty surrounding the PI mass limit, we consider three cases: (i) $m_1 < 65\,M_{\odot}$, (ii) $m_1 < 50\,M_{\odot}$, and (iii) $m_1 < 40\,M_{\odot}$.
For brevity, we abbreviate each model and add a subscript indicating the primary mass cutoff, e.g. BPL$_{65}$ indicates the \textsc{Broken Power Law} model with a primary mass cutoff of $m_1 < 65 \, M_{\odot}$.

\subsection{Injections}
\label{sec:injections}

An essential part of this work is that we add injections directly to GW detector strain data. This allows us to compare our detected BBH events against the population models, accounting for selection bias.
We use the SEOBNRv4 waveform approximant~\cite{SEOBNRv4} and inject simulated waveforms uniformly in time throughout the O1-O3a observing runs.
In the case of O1 and O2, we use the public GW strain data available through the Gravitational Wave Open Science Center (GWOSC)~\cite{gwosc}.
At the time of this publication, the GW strain data for the complete O3a observing run was not publicly available. Instead, we used the segments of data available from the GWTC-2 GWOSC data release~\cite{gwosc} to emulate the O3a sensitivity, and we repeated injection analysis on these data segments until realizing the full O3a observation time.

The component masses for each simulated BBH event are drawn from the tested stellar-mass model under consideration (see Section~\ref{sec:models}). Component black hole spins are drawn from a uniform aligned spin distribution.
The binary source distance is drawn assuming uniform density in co-moving volume up to redshift $z=2$~\cite{Planck2015}.
Other extrinsic binary parameters including sky location and inclination angle are randomly selected.

We recover the simulated BBH events using the cWB pipeline and select events with a FAR less than $1\,\mathrm{yr}^{-1}$.
The right plot in Figure~\ref{fig:model_det_m1} displays the distribution of detected events for each population model. By comparison to the astrophysical distributions, the detected distributions favor more massive BBH mergers due to selection effects caused by the ground-based detectors and the cWB search pipeline.

\renewcommand{\arraystretch}{1.4}
\begin{table}[hbt]
\begin{tabular}{lllll}
\hline
\hline
 Event & $\mathrm{SNR}_{\mathrm{net}}$ & $f_0\,$[Hz] & $M_{\mathrm{cWB}}\, [M_{\odot}]$ & $M_{\mathrm{GWTC}}\, [M_{\odot}]$\\
\hline
        GW190521 &    14.4 &        57.8 &               193.2 &  $157.9_{-20.9}^{+37.5}$ \\
 GW190706\_222641 &    12.7 &        74.1 &               139.2 &  $101.1_{-13.5}^{+18.0}$ \\
 GW190602\_175927 &    11.1 &        76.9 &               126.8 &  $114.0_{-15.5}^{+18.4}$ \\
 GW190519\_153544 &    14.0 &        89.2 &               115.5 &  $104.2_{-15.0}^{+14.5}$ \\
        GW170729 &    10.2 &        81.8 &               110.5 &   $84.4_{-11.1}^{+15.8}$ \\
 GW190701\_203306 &    10.2 &        93.3 &               104.3 &    $94.1_{-9.2}^{+11.6}$ \\
 GW190521\_074359 &    24.7 &        91.4 &                92.9 &     $74.4_{-2.4}^{+6.9}$ \\
        GW170814 &    17.2 &       115.4 &                90.2 &     $55.9_{-2.6}^{+3.4}$ \\
 GW190421\_213856 &     9.3 &        87.2 &                84.6 &    $71.7_{-8.6}^{+12.5}$ \\
 GW190727\_060333 &    11.4 &       103.5 &                81.2 &    $65.8_{-7.3}^{+10.8}$ \\
 GW190503\_185404 &    11.5 &       109.8 &                78.8 &     $71.3_{-8.1}^{+9.3}$ \\
        GW170823 &    10.8 &       114.2 &                73.5 &    $68.7_{-8.1}^{+10.8}$ \\
        GW150914 &    25.2 &       118.7 &                68.6 &     $66.1_{-3.3}^{+3.8}$ \\
 GW190915\_235702 &    12.3 &       119.9 &                67.1 &     $59.5_{-6.2}^{+7.6}$ \\
        GW170104 &    13.0 &       147.0 &                65.3 &     $51.0_{-4.1}^{+5.3}$ \\
 GW190828\_063405 &    16.6 &       103.5 &                60.9 &     $57.5_{-4.5}^{+7.4}$ \\
 GW190408\_181802 &    14.8 &       117.3 &                52.2 &     $42.9_{-2.9}^{+4.0}$ \\
        GW151226 &    11.9 &       120.6 &                45.8 &     $21.5_{-1.5}^{+6.2}$ \\
        GW190412 &    19.7 &       108.5 &                45.6 &     $38.4_{-3.7}^{+3.8}$ \\
 GW190517\_055101 &    10.7 &       133.4 &                41.8 &    $61.9_{-9.6}^{+10.1}$ \\
 GW190512\_180714 &    10.7 &       125.7 &                33.1 &     $35.7_{-3.4}^{+3.8}$ \\
        GW170608 &    14.1 &       116.9 &                17.4 &     $18.6_{-0.7}^{+3.2}$ \\

\hline
\hline
\end{tabular}
\caption{List of 22 BBH merger event candidates observed by cWB during the O1, O2, and O3a observing runs.
Summary statistics, including the central frequency $f_0$ and network signal-to-noise ratio $\mathrm{SNR}_{\mathrm{net}}$, are reconstructed by the cWB pipeline and fed into XGBoost to estimate the source frame total mass \Mcwb (described in Section~\ref{sec:ts}). Events in the table are ordered according to \Mcwb. The estimated source frame mass \Mcwb can be compared against the estimated source frame mass $M_{\mathrm{GWTC}}$ reported in the LIGO-Virgo Transient Catalogs~\cite{GWTC1, GWTC2}.
}
\label{tab:events}
\end{table}

\subsection{Population outlier test statistic}
\label{sec:ts}

In this study, we require a test statistic that captures the characteristics of an outlier to the stellar-mass BBH population.
One logical statistic to consider is the source frame primary mass $m_1$, since stellar-mass BBH events are not expected to have $m_1$ inside the PI mass gap.
Ideally, one could use the source frame primary mass $m_1$ estimated via Bayesian inference.
However, since we inject and recover $\mathcal{O}(100,000)$ simulated BBH events in this study, this is not a feasible approach.

\begin{figure}[!tbh]
    \centering
    \includegraphics[width=0.48\textwidth]{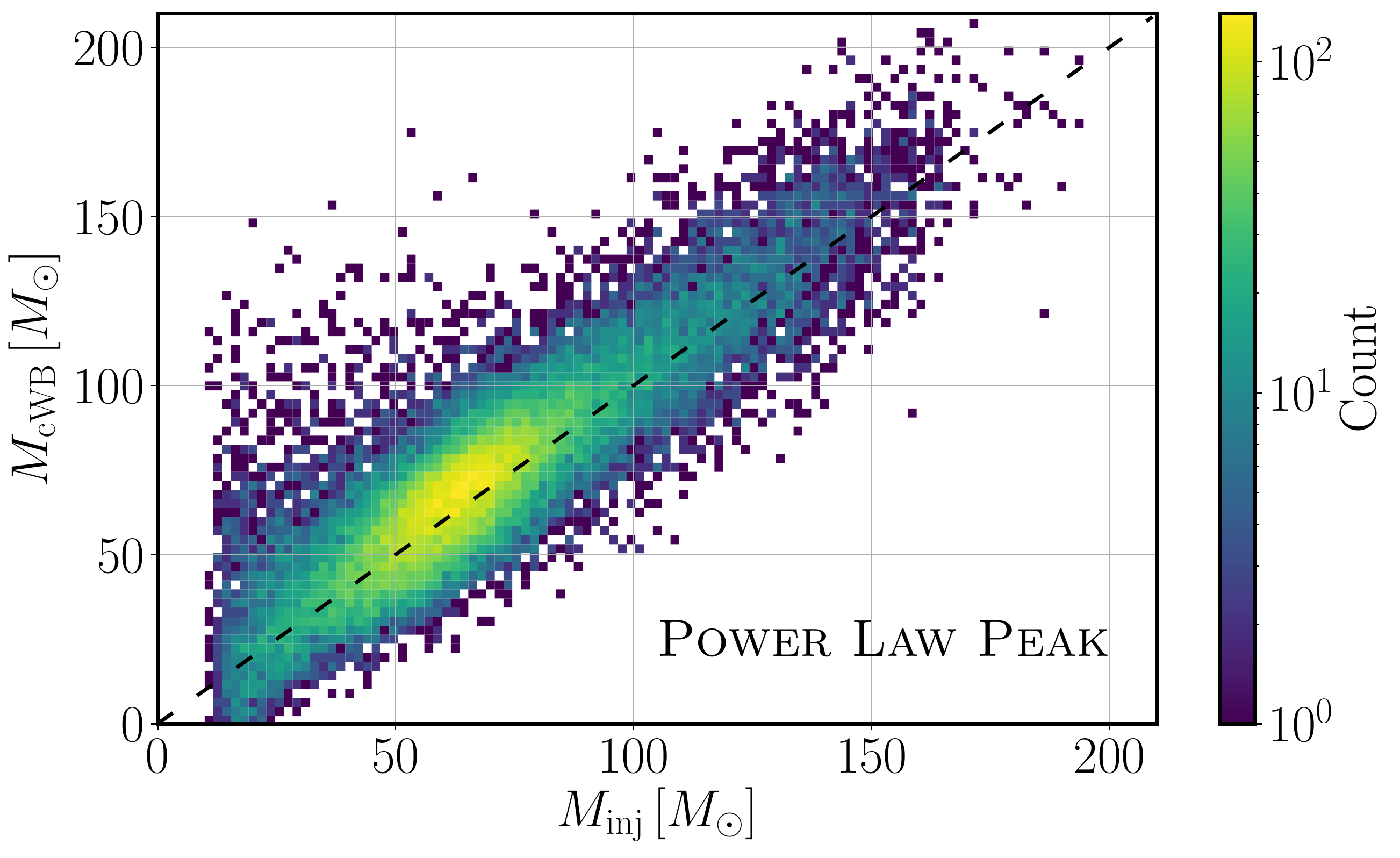}
    \caption{Estimated source frame total mass \Mcwb and injected source frame total mass $M_{\mathrm{inj}}$ for simulated BBH events drawn from the \textsc{Power Law  + Peak} mass distribution model. Dashed black line indicates the identity line.}
    \label{fig:rec_inj}
\end{figure}

Instead, we utilize cWB in combination with a supervised machine learning (ML) algorithm XGBoost~\cite{XGBoost} to rapidly estimate the source frame total mass for a given event.
This estimate, which we call \Mcwb, serves as our test statistic.
The injected source frame total mass $M_{\mathrm{inj}}$ is designated as the target variable in our ML regression model, and the summary statistics reconstructed by cWB are used as input features to estimate the target variable.
Figure~\ref{fig:rec_inj} shows the injected $M_{\mathrm{inj}}$ and estimated \Mcwb for simulated BBH events drawn from the \textsc{Power Law  + Peak} mass model.
In Appendix C.4 of Ref.~\cite{POP_GWTC2}, we use a simplified approach without ML, but we expect the method presented in this paper to be more sensitive to measure outliers of the stellar-mass BBH population.

To generate a robust and unbiased estimate of the source frame total mass, we train our ML algorithm on a simulation set independent from the tested stellar-mass population simulation sets.
In this case, we use the IMRPhenomPv2 waveform approximant~\cite{IMRPhenomPv2} and inject simulated waveforms uniformly throughout O1, O2, and our emulated O3a observation run.
For each training event, the primary mass \Mone is drawn from a uniform distribution \Mone $\in [1,150] \, M_\odot$, and the mass ratio $q$ is fixed equal to one.
The component spins are drawn from an isotropic distribution.
Extrinsic parameters including sky location, orientation, and distance are each drawn from uniform distributions.
We again consider only events detected by cWB with FAR less than $1\,\mathrm{yr}^{-1}$.
More information regarding our ML training and tuning procedure can be found in Appendix~\ref{sec:training}.

\subsection{Statistical procedure}
\label{sec:procedure}

Our statistical procedure is divided into two steps.
First we identify which events, if any, in our sample set are potential outliers to the stellar-mass BBH population.
And second, we remove any potential outliers and determine if the remaining distribution of events is consistent with the stellar-mass BBH population. We use \Mcwb, defined in Section~\ref{sec:ts}, as an outlier test statistic where a higher value of \Mcwb indicates a greater probability for the event to have originated inside the PI mass gap.

First, to estimate the significance of potential outliers, we consider a bootstrap resampling procedure. In this procedure, we identify the event $x$ which has the highest value of \Mcwb from our sample set of $N$ event candidates.
We then randomly draw $N$ simulated events with replacement from the model distribution and select the simulated event from the set which has the highest value of \Mcwb, i.e., we select the outlier from the simulated set.
We repeat this routine 10,000 times, and thus estimate the distribution of outliers events $O$ predicted by the model.
So, the probability for a model to produce an outlier as significant as event $x$ is given as:
\begin{equation}
\label{eq:1}
\mathrm{Pr}(M_{\mathrm{cWB}}^x \geq M_{\mathrm{cWB}}^O \, | \, H_0 ) \, ,
\end{equation}
where $H_0$ represents the null hypothesis that event $x$ is consistent with the model outlier distribution $O$. This is the one-sided (right tail) p-value.
If this probability is less than our conservatively chosen threshold $\alpha = 0.01$, we claim that event $x$ is a potential outlier to the tested population model.

In the case where event $x$ is found to be a potential outlier, we omit the event from the sample set and reanalyze the remaining $N-1$ events.
We repeat this process until all outliers are identified.

Once we have identified the number of population outliers $N_{\mathrm{out}}$, we then study whether the remaining sample set of $N - N_{\mathrm{out}}$ events are consistent with the population model.
We again consider a bootstrap resampling procedure where the BBH events in our sample set are ordered according to \Mcwb.
We then randomly draw 10,000 batches of $N - N_{\mathrm{out}}$ simulated events and sort each batch according to \Mcwb.
The $n^{\mathrm{th}}$ ranked event from our sample set is compared against the $n^{\mathrm{th}}$ ranked simulated event from each batch.
Thus, we can identify if the $n^{\mathrm{th}}$ ranked event in our sample set is consistent with predictions made by the tested population model.

To further quantify the consistency between the sample distribution and tested population model, we apply the Anderson-Darling (AD) goodness-of-fit test~\cite{AD_test:1952}.
Compared to the Kolmogorov-Smirnov (KS) test~\cite{kolmogorov:1933, smirnov:1948}, the AD test is expected to be more sensitive at the tails of the distribution and so is more suitable for this analysis.
If the AD p-value $p_{\mathrm{AD}}$ is less than $\beta = 0.01$, then we find the sample distribution is inconsistent with the tested model distribution. 

\section{Results} \label{sec:results}

\subsection{PI mass limit of $65\, M_\odot$}
\label{sec:65}

In this section, we present the results of our analysis under three separate assumptions of the PI mass limit. First, we examine stellar-mass BBH models under the assumption of a $65\, M_\odot$ PI mass limit: BPL$_{65}$ and PLP$_{65}$, defined in Section~\ref{sec:models}.
Our sample distribution consists of 22 BBH events detected by cWB (see Table~\ref{tab:events}).
We order these events according to \Mcwb. Here, the first rank event is GW190521 with $M_\mathrm{cWB} = 193.2\, M_\odot$.

\begin{figure}[hbt]
    \centering
    \includegraphics[width=0.48\textwidth]{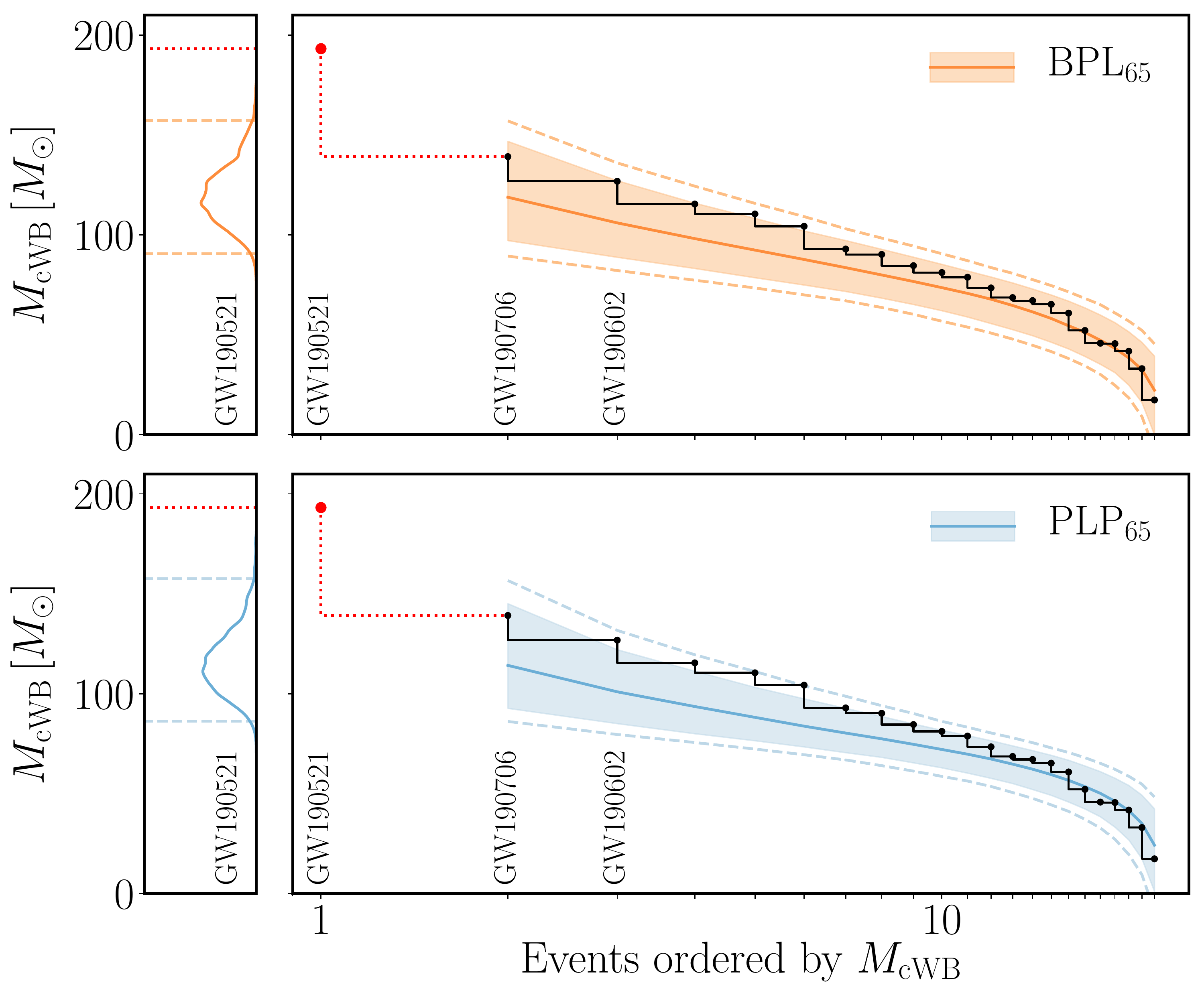}
    \caption{Distribution of BBH events detected by cWB, shown in red/black, ranked according to \Mcwb and compared to predictions made by stellar-mass models BPL$_{65}$ (top) and PLP$_{65}$ (bottom), which assume a PI mass limit of $65\, M_\odot$. Event names are abbreviated. Shaded region indicates 90\% confidence interval, dashed lines indicate 98\% confidence interval. GW190521 is identified as an outlier to both population models. The remaining BBH events are consistent with both models.}
    \label{fig:pval_65}
\end{figure}

We compare GW190521 against outliers predicted by each model to determine if the event is a stellar-mass population outlier.
Specifically, we generate batches of 22 simulated events from the \bpli and \plpi models and compare GW190521 against the outlier from each batch.
The rotated histograms in Figure~\ref{fig:pval_65} show \Mcwb for GW190521 (red dotted line) compared against the distribution of \Mcwb for outliers predicted by \bpli (orange) and \plpi (blue). The dashed lines in each histogram denote the 98\% confidence predicted by the model.
Using Equation~\ref{eq:1}, we find the probability for the \bpli and \plpi models to produce an outlier as significant as GW190521 to be less than $0.0001$. 
So, assuming the PI mass limit exists as a sharp cutoff at $65\, M_\odot$, we find GW190521 to be an outlier to the stellar-mass population.

Next, we investigate if any other events are outliers to the stellar-mass population. 
We omit GW190521, and re-analyze the remaining 21 detected BBH events in our sample set against batches of 21 simulated BBH events predicted by each model.
The highest ranked event in this sample set is GW190706\_222641 with $M_\mathrm{cWB} = 139.2\, M_\odot$.
After applying the same statistical procedure as above, we calculate a p-value of 0.109 and 0.082 for the \bpli and \plpi model, respectively, and so this event is consistent with both population models.

After identifying all potential population outliers, in this case just GW190521, we examine whether the remaining sample distribution is consistent with the tested population models.
Figure~\ref{fig:pval_65} shows the estimated \Mcwb for our sample set (black) compared to predictions made by the \bpli (orange) and \plpi (blue) models.
The dark orange (blue) line represents the median value of \Mcwb predicted by the \bpli (\plpi) model, the shaded region indicates 90\% confidence interval, and the dashed lines indicate 98\% confidence interval.
We find that all 21 events lie within the 98\% confidence interval predicted by each model. Applying the AD goodness-of-fit test, we estimate $p_{\mathrm{AD}}$ for the \bpli model and \plpi model to be greater than 0.25.
So, the distribution of detected events, excluding GW190521, is consistent with both stellar-mass BBH models.

\subsection{PI mass limit of $50\, M_\odot$}
\label{sec:50}

Next, we repeat our statistical analysis for stellar-mass BBH models assuming a PI mass limit of $50\, M_\odot$: \bplii and \plpii.
Here, we again compare the \Mcwb distribution of the 22 BBH events detected cWB (our sample distribution) against the \Mcwb distribution predicted by the two models.

\begin{figure}[hbt]
    \centering
    \includegraphics[width=0.48\textwidth]{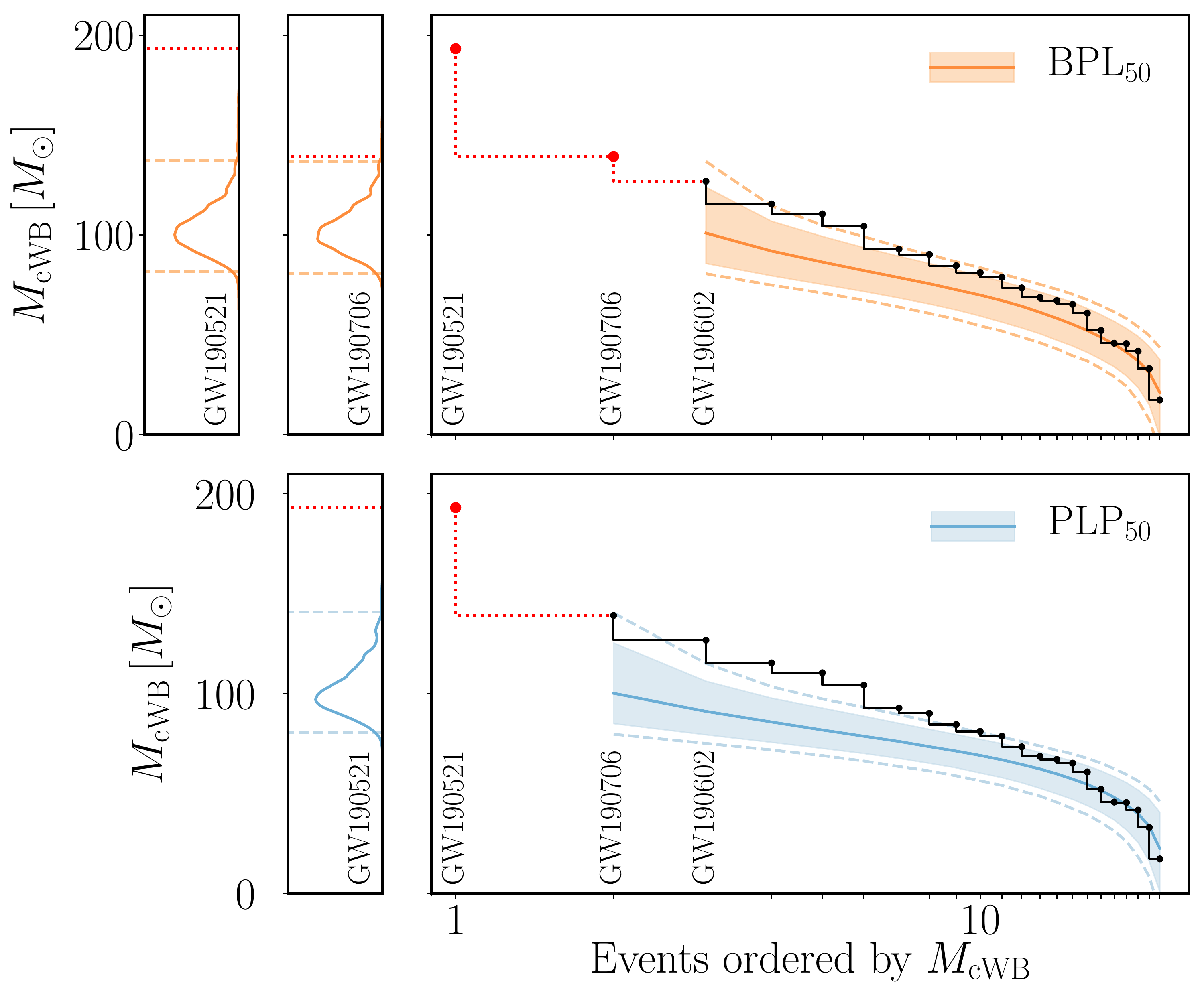}
    \caption{Distribution of BBH events detected by cWB, shown in red/black, compared to predictions made by stellar-mass models BPL$_{50}$ (top) and PLP$_{50}$ (bottom), which assume a PI mass limit of $50\, M_\odot$. For the BPL$_{50}$ model, GW190521 and GW190706\_222641 are evaluated as potential outliers, and the remaining events are consistent with the population model. For the PLP$_{50}$ model, only GW190521 is identified as an outlier, while the remaining distribution of events are inconsistent with the model.}
    \label{fig:pval_50}
\end{figure}

As in Section~\ref{sec:65}, we find GW190521 to be an outlier to the tested stellar-mass population models (p-value $< 0.0001$).
So, we re-analyze the remaining 21 BBH detections. 
Here, the highest ranked event from this set, GW190706\_222641, has an outlier p-value of 0.009 when compared to the \bplii model and an outlier p-value of 0.011 when compared to the \plpii model. 
By definition of our \textit{a priori} p-value threshold $\alpha = 0.01$, we treat this event as a potential outlier to the \bplii model but not the \plpii model. No other potential outliers are identified for the \bplii stellar-mass model.
In Figure~\ref{fig:pval_50}, the rotated histograms show the estimated \Mcwb of potential outliers compared to predictions made by each population model. Here, it is evident that GW190521 is a clear outlier to both models, whereas GW190706\_222641 lies just outside the 98\% confidence interval for the \bplii model.

Next, we examine the remaining distribution of detected events. The main panels in figure~\ref{fig:pval_50} show this distribution compared to predictions made by each model. Here, we observe that the sample distribution appears to deviate from the predicted distributions, particularly the \plpii model distribution. That is, there are more high mass events in the sample distribution compared to the predictions made by each model.
To quantify this deviation, we apply the AD test. For the \bplii stellar-mass model, we analyze the 20 non-outlier BBH detections and calculate $p_{\mathrm{AD}}$ to be 0.039 -- so the detected events are consistent with the model.
For the \plpii model, we analyze the 21 non-outlier BBH detections, including GW190706\_222641, and calculate $p_{\mathrm{AD}}$ to be 0.006. This indicates that the sample distribution is inconsistent with the \plpii model. As a cross-check, we examine the case of evaluating GW190706\_222641 as a potential outlier, since it barely passes our outlier p-value threshold. Applying the AD test for this case, we calculate $p_{\mathrm{AD}}$ to be 0.036. So, if GW190706\_222641 is treated as a potential outlier, then the remaining sample distribution is consistent with \plpii. 

\subsection{PI mass limit of $40\, M_\odot$}
\label{sec:40}

\begin{figure}[!hbt]
    \centering
    \includegraphics[width=0.48\textwidth]{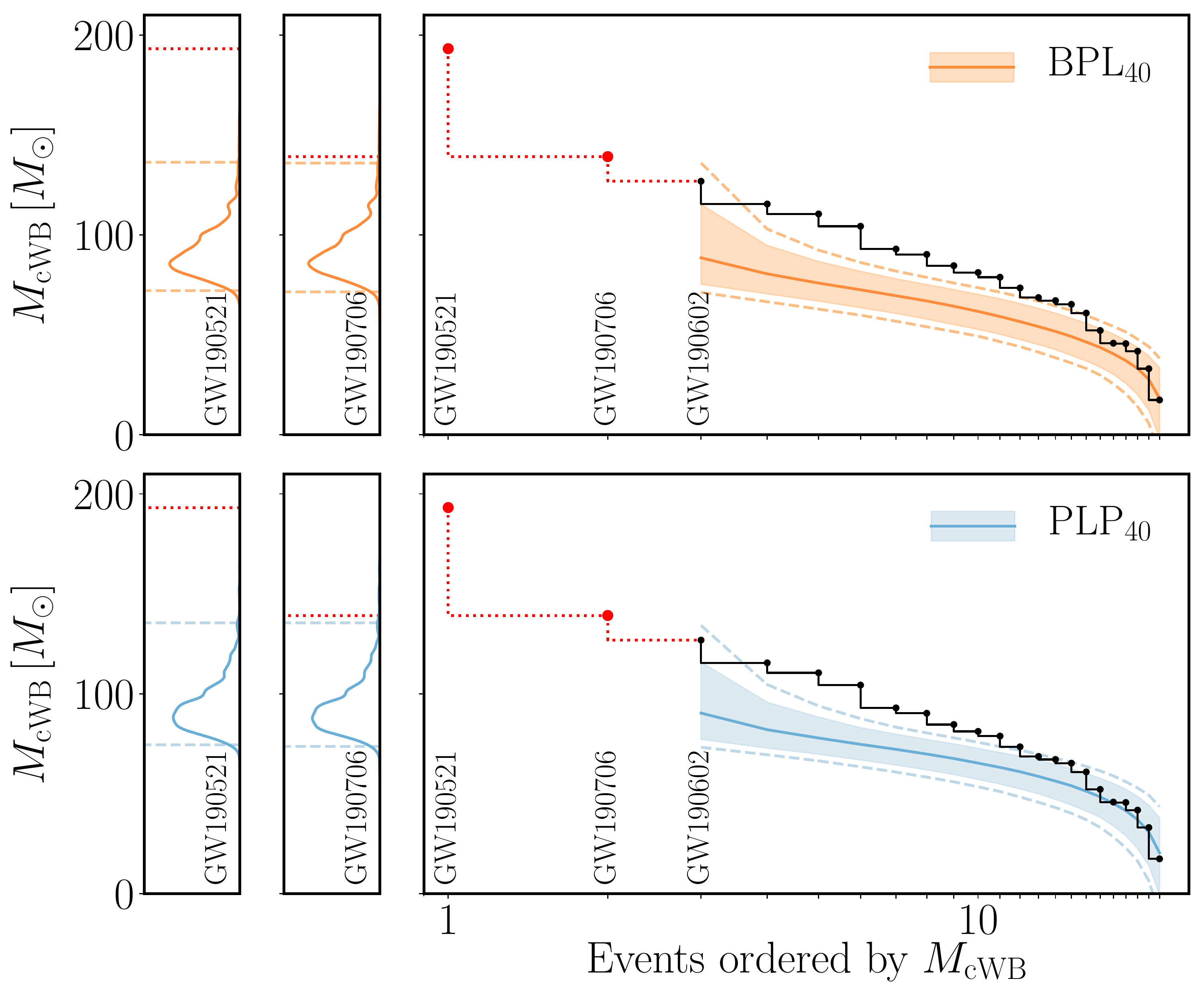}
    \caption{Distribution of BBH events detected by cWB, shown in red/black, compared to predictions made by stellar-mass models BPL$_{40}$ (top) and PLP$_{40}$ (bottom). Two events --- GW190521 and GW190706\_222641 --- are found to be potential population outliers. The remaining 20 BBH events are inconsistent with both stellar-mass population models.}
    \label{fig:pval_40}
\end{figure}

In the final case, we consider a PI mass limit of $40\, M_\odot$. Our results for this case are summarized in Figure~\ref{fig:pval_40}.
As in the previous two cases, we first compare GW190521 against the distribution of outliers predicted by each model. We again find GW190521 to be an outlier with a p-value less than $0.0001$ for both models (leftmost rotated histogram). We then re-analyze the remaining 21 BBH events in our sample set to test if the highest ranked event in this set, GW190706\_222641, is an outlier to the stellar-mass BBH models. For this event, we calculate a p-value of 0.007 and 0.006 for the \bpliii and \plpiii models, respectively (second rotated histogram). So, this event is also classified as a potential outlier.
We then re-run our analysis for the remaining 20 detected BBH events and find no additional outliers in the sample set.

The main panels in Figure~\ref{fig:pval_40} show the sample distribution, according to \Mcwb, compared to predictions made by each stellar-mass population model. Here, we notice that more high mass events are found compared to the predictions made by either population model.
Applying the AD test, we find $p_{\mathrm{AD}}$ for both the \bpliii model and the \plpiii model to be less than 0.001. So, if the PI mass limit is at $40\, M_\odot$, then we have two potential outliers -- GW190521 and GW190706\_222641 -- and out of the remaining 20 BBH events, there are more high mass events than predicted by the models.

\section{Conclusions}
\label{sec:conclusions}

In this paper, we introduce a method to determine if the distribution of BBH events detected by cWB is consistent with the stellar-mass population.
Detector noise is unavoidable and could affect the inference and understanding of any given event, thereby making an otherwise ordinary BBH merger appear as an outlier.
Our analysis operates directly on search statistics by detecting simulated waveforms representative of the stellar-mass population injected into the detector strain data.
This way, any inference errors caused by the detector noise are captured within the model distribution.

We began this study to further investigate the origin of the GW190521 signal and to understand whether an event like GW190521 could be imitated by a stellar-mass BBH merger. 
We conclude that if the PI mass gap begins at or below $65\,M_\odot$, then GW190521 is a clear outlier to the stellar-mass BBH population and that its outlier significance cannot be explained by statistical fluctuations.
In general, any population study, such as the analysis presented in this paper, is dependent on the sample distribution. So, we expect the significance of stellar-mass population outliers presented here to fluctuate following future observing runs.
However, it is critical to note that we do not expect the outlier significance of GW190521 to change in a meaningful way (See Appendix~\ref{sec:further_investigation}).

Turning our attention to the remaining distribution of 21 BBH events detected by cWB, we find it unlikely that this sample distribution is consistent with the stellar-mass BBH population, assuming the PI mass limit exists at $50\,M_\odot$.
In this case, there is also some evidence that GW190706\_222641 could be an outlier to the stellar-mass population. If the PI mass limit is at $40\,M_\odot$, differences between the sample distribution and stellar-mass population models cannot be reconciled. In this case, there is overwhelming evidence that cWB has detected more high mass BBH events compared to predictions made by the stellar-mass models.

In future work, we aim to expand upon this analysis by testing BBH population models corresponding to specific astrophysical channels.
In this case, the inclusion of spin effects into our test statistic will become important to distinguish between different alternative scenarios.

\begin{acknowledgments}

This work was supported by NSF Grant No.
PHY 1806165.
We gratefully acknowledge the computational resources provided by LIGO-Virgo. This material is based upon work supported by NSF’s LIGO Laboratory which is a major facility fully funded by the National Science Foundation.
This research has made use of data, software and/or
web tools obtained from the Gravitational Wave Open
Science Center, a service of LIGO Laboratory, the LIGO
Scientific Collaboration and the Virgo Collaboration.
The authors thank Shanika Galaudage for providing mass samples corresponding to the GWTC-2 population models.
I. B. acknowledges support by the National Science Foundation under grant No. PHY 1911796, the Alfred P. Sloan Foundation and by the University of
Florida.
We thank Ik Siong Heng, Erik Katsavounidis, Peter Shawhan, and Michele Zanolin for their continued participation and effort in reviewing cWB analyses.
In this analysis, we make use of open source Python packages including \textsc{NumPy}~\cite{numpy}, \textsc{SciPy}~\cite{scipy}, \textsc{Pandas}~\cite{pandas}, \textsc{Matplotlib}~\cite{matplotlib}, \textsc{Seaborn}~\cite{seaborn}, and \textsc{scikit-learn}~\cite{scikit-learn}.

\end{acknowledgments}

\appendix

\section{Consistency check on phenomenological population models}

This section serves as a consistency check on two phenomenological population models, \textsc{Broken Power Law} and \textsc{Power Law  + Peak}, which are fit to the population of BBH mergers detected by LIGO-Virgo.
This is an update to the work presented in Ref.~\cite{POP_GWTC2}, Appendix C.4.

\begin{figure}[hbt]
    \centering
    \includegraphics[width=0.48\textwidth]{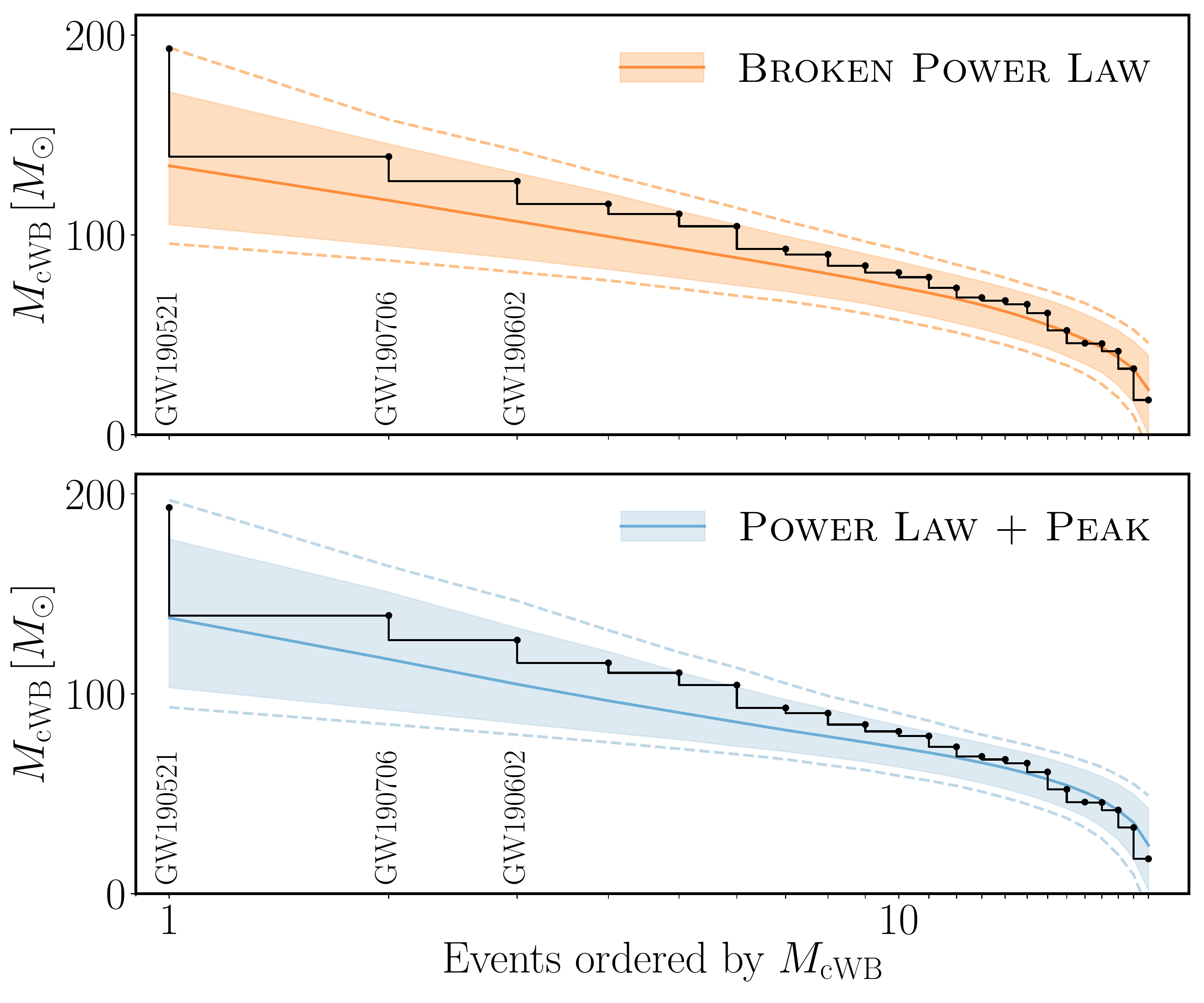}
    \caption{Distribution of BBH events detected by cWB, shown in black, ranked according to \Mcwb and compared to predictions made by the \textsc{Broken Power Law} model (top) and \textsc{Power Law + Peak} model (bottom). Shaded region indicates 90\% confidence interval, dashed lines indicate 98\% confidence interval. The BBH events detected by cWB are consistent with both population models.}
    \label{fig:pval_full}
\end{figure}

Here, we apply the same methodology presented in Section~\ref{sec:method}.
First, we compare GW190521 against the distribution of outliers predicted by each model.
We estimate the probability for the \textsc{Broken Power Law} model and the \textsc{Power Law + Peak} model to produce an outlier as significant as GW190521 to be 0.0104 and 0.015, respectively.
This is consistent with the result reported in Ref.~\cite{POP_GWTC2}, where we estimate the outlier p-value of GW190521 to be 0.053 for the \textsc{Broken Power Law} model and 0.077 for the \textsc{Power Law + Peak} model.
Here, we expect the change in significance is due to the improved sensitivity to population outliers with the updated method presented in this paper.
The results of this test are summarized in Figure~\ref{fig:pval_full}.
Analyzing the full distribution of events detected by cWB, we find that the sample distribution is consistent with both population models ($p_{\mathrm{AD}} > 0.25$).
Overall, we cannot make any conclusive statements regarding which model fits the observed data better. Future BBH detections could help resolve the differences between the models and point to a preferred model.

\section{Further investigation of the outlier significance of GW190521} \label{sec:further_investigation}

One main conclusion of this work is that GW190521 is inconsistent with the stellar-mass population, assuming a PI mass limit at or below $65\, M_\odot$.
This result, however, is inherently dependent on our sample size and is subject to change following additional detections in future observing runs.
Nevertheless, we expect the outlier significance of GW190521 presented here will not meaningfully change following future observing runs.

To test this, we consider a hypothetical scenario where the cWB pipeline identified 100 BBH events during the O1, O2, and O3a observing runs (instead of 22 BBH events). Under this scenario, we conservatively assume that GW190521 persists as the first rank event, according to \Mcwb.
Following the statistical procedure described in Section~\ref{sec:procedure}, we estimate the outlier p-value of GW190521 to be less than $0.0001$.
Though this p-value is admittedly limited by our simulated sample size, our conclusion remains that no simulated BBH event representative of the stellar-mass population is as significant as GW190521.

\section{ML training feature list}
\label{sec:features}

We utilize XGBoost, a boosted decision tree ML algorithm~\cite{XGBoost}, as a regression tool to estimate the source frame total mass \Mcwb for a given event.
To estimate the target variable, ML algorithms require input features which are correlated with the target variable. In our case, the target variable is the source frame total mass $M_\mathrm{inj}$, and the ML features are summary statistics reconstructed by cWB.
Below is the list of 13 summary statistics used in the training process, ordered according to the \textit{gain} importance metric. 
Unsurprisingly, the most important feature is a measure of the central frequency $f_0$, which is expected to be inversely proportional to the detector frame total mass.
Further attempts at feature pruning reduced our evaluation score on a validation data set.
Some summary statistic definitions below are equivalent to those found in Ref.~\cite{CWB_ML}.

\begin{itemize}
  \item $f_0$ --- Energy weighted signal central frequency.
  \item $\Delta F_\mathrm{s}$ --- Energy weighted signal bandwidth.
  \item $O_\mathrm{n}$ --- Observation run number. Including the observation run number allows the ML algorithm to respond to changes in detector sensitivity across separate observing runs.
  \item $n_\mathrm{f}$ --- Effective number of time-frequency resolutions used for event detection and waveform reconstruction.
  \item $\Delta T_\mathrm{s}$ --- Energy weighted signal duration.
  \item $F_\mathrm{M}$ --- Chirp mass energy fraction, chirp mass goodness of fit metric, defined in Ref.~\cite{Tiwari:2015}. 
  \item $\mathcal{M}$ --- Chirp mass parameter estimated in the time-frequency domain, defined in Ref.~\cite{Tiwari:2015}.
  \item SNR$_\mathrm{net}$ --- Quadrature sum of the reconstructed signal SNR found in each GW detector.
  \item $Q_\mathrm{0}$ --- An estimation of the effective number of cycles in a cWB event~\cite{Qveto}. 
  \item $e_\mathrm{M}$ --- Chirp mass ellipicity, chirp mass goodness of fit metric, defined in Ref.~\cite{Tiwari:2015}. 
  \item $c_\mathrm{c}$ --- Coherent energy divided by the sum of coherent energy and null energy, defined in Ref.~\cite{Klimenko:2008fu}.  
  \item $Q_\mathrm{1}$ --- The waveform shape parameter~\cite{Qveto} developed to identify a characteristic family of (blip) glitches present in the detectors~\cite{GWTC1, McIver2012}.
  \item $E_\mathrm{c}/L$ --- Ratio of the coherent energy to the network likelihood.
\end{itemize}

\section{ML training and tuning}
\label{sec:training}

Our training set consists of simulated BBH events recovered by cWB, described in section~\ref{sec:ts}.
We use the Huber loss function~\cite{huber_loss} as the learning objective for our XGBoost regression model. This loss function is expected to be less sensitive to outliers compared to the squared loss function.
The \texttt{base\_score}, which represents the initial prediction for all instances, is initialized to the mean injected source frame total mass of the training data set $\bar{M}_\mathrm{inj}$.
To optimize the number of decision trees, we employ early stopping which restricts the number of trees based on the model performance on an independent validation dataset.
Here, early stopping effectively limits over-fitting.
To tune the remaining XGBoost hyper-parameters, we perform a grid search over 540 combinations of hyper-parameters.
Here, we use 5-fold cross validation to further prevent over-fitting. Selected hyper-parameters are shown in bold in Table~\ref{tab:tuning}.

\renewcommand{\arraystretch}{1.4}
\begin{table}[hbt]
\begin{tabular}{lc}
\hline
\hline
hyperparameter & entry\\
\hline
\texttt{objective} & \texttt{reg:pseudohubererror}\\
\texttt{tree\_method} & \texttt{gpu\_hist}\\
\texttt{grow\_policy} & \texttt{lossguide}\\
\texttt{base\_score} & $\bar{M}_\mathrm{inj}$\\
\texttt{n\_estimator}s & 20000$\dagger$\\
\texttt{learning\_rate} & \textbf{0.03}, 0.1\\
\texttt{max\_depth} & \textbf{5}, 7 , 9, 11, 13\\
\texttt{min\_child\_weigh}t & 1, \textbf{5}, 10\\
\texttt{gamma} & \textbf{1.0}, 2.0, 5.0\\
\texttt{colsample\_bytree} & 0.6, 0.8, \textbf{1.0}\\
\texttt{subsample} & 0.6, \textbf{0.8}\\
\hline
\hline
\end{tabular}
\caption{List of tuned hyper-parameters. Bold parameters indicate hyper-parameters selected through tuning with 5-fold cross validation. The \texttt{base\_score} is initialized to the mean source frame total mass of the training data set $\bar{M}_\mathrm{inj}$ $\dagger$: the \texttt{n\_estimators} hyper-parameter is optimized using early stopping, described in the main text.
}
\label{tab:tuning}
\end{table}

\newpage
\bibliography{main.bib}

\end{document}